\documentclass[10pt]{article}
\usepackage{geometry}                
\usepackage{graphicx}
\usepackage{amssymb}
\usepackage{url}
\usepackage{subfigure}
\usepackage{makeidx}

\title{Simple fluidic digital half-adder}
\author{Alex J. L. Morgan\textsuperscript{a},
David A. Barrow\textsuperscript{a},\\
Andrew Adamatzky\textsuperscript{b} and 
Martin M. Hanczyc\textsuperscript{c}}

\begin{document}
\maketitle

\centerline{\textsuperscript{a}~Cardiff School of Engineering, Cardiff University, Cardiff, Wales, UK}

\centerline{\textsuperscript{b}~Unconventional Computing Centre, UWE, Bristol, UK}

\centerline{\textsuperscript{c}~Laboratory of Artificial Biology, Centre for Integrative Biology, University of Trento, Italy}

\vspace{1cm}

\begin{abstract}
A fluidic one-bit half-adder is made of five channels which intersect at a junction. Two channels are inputs, two channels are outputs and one channel is the drain. The channels direct fluid from input fragments to output fragments and the streams of fluid interact at the junctions. Binary signals are represented by water droplets introduced in the input channels: presence of a droplet in an input or output segments symbolises logical {\sc True}, absence --- {\sc False}.  The droplets travel along channels by following a path of least resistance unless deflected at the junction. We demonstrate the function of the half-adder in both computer modelling and laboratory experiments, and propose a design of a one-bit full adder based on simulation.
\end{abstract}

\section{Introduction}

Microfluidics provides an ideal environment to precisely control the input parameters of flowing liquids.  In this way one can work out in detail the experimental conditions required for the implementation of many different preferred outcomes, including computational logic gates as explored here.  Once the "rules" for computational implementation have been explored in controlled fluidic environments, the droplet technology can then be developed in more complex and real world environments. The implementation of computation with input parameters coming from the natural environment may have impact on future theranostic applications where programmed delivery and controlled release of targeted chemistries would provide a great advance in medical therapies.  

Fluids have been used as information carriers and computing substrates since the early 1960s, and this quickly developed into a strong field of science and engineering including a range of high-profile conferences~\cite{pavlin1965proceedings, stephens1968proceedings}. The advantages of fluidic computing and actuating devices are reliability, tolerances to shock, vibration, radiation, as well as low costs~\cite{conway1971guide, kleinstreuer2009modern}. Fluidic devices, and their modern reincarnations --- microfluidic devices, have disseminated through the applied sciences and engineering: aircraft control and actuation, stimuli-responsive gels, cell biology, hydraulics, colloid science, lab-on-a-chip and organ-on-a-chip, enzymatic reactions, non-silicon memory devices, capillary  electrophoresis, fuel cells, DNA microarrays, nano-technology, high-throughput biochemical assays~\cite{madou2011solid, ho2012design, erickson2004integrated, murthy2009materials, oosterbroek2003lab, gomez2008biological, edel2009nanofluidics, feldstein1999array, kumar2010microfluidic}. Fluidic devices also found an application in photonics  as optofluidic devices and sub-systems~\cite{yeshaiahu2010optofluidics}: self-assembly techniques and micro fabrication, synthesis of photonics crystals, and optical lab-on-a-chip devices.

Computational fluidic devices must be controlled either by an external computer or through embedded control units. The former is inefficient and the latter requires structural, or physical, programming. The structural programming means encoding of logical values into intensities of fluid pressure or velocity and implementation of logical operations and binary arithmetical circuits at the cascaded junctions where the streams collide. In first five years, since the first ever fluidic logic device was created at the Harry Diamond Lab~\cite{conway1971guide}, a number of fluid-based logic elements have been designed. The early logic gates were based on jet/stream attachment, e.g. threshold logic devices~\cite{florea1973fluidic}, 
and comprised of complex systems of channels with variable width, branching and expansion outlets~\cite{keefe1974discovery, healey1973fluidic, volcik1975fluidic, mitchusson1963fluid}. With the resurrection of fluidics in the late 1990s, a range of non-trivial devices were implemented including devices for sequential routing of a certain number of droplets to a specified location~\cite{um2009microfluidic}, and addition of droplet volumes via fusion,~\cite{um2008microfluidic}, logic via bubbles in microfluidic channels~\cite{prakash2007microfluidic}, memory devices with viscous polymer fluids~\cite{groisman2003microfluidic}, and fluidic transistors that explore the relation between a channel and the electrostatic interactions within the channel~\cite{karnik2006field}.

In parallel, and apparently independent to fluidics, the field of collision-based computing~\cite{CBC}  emerged in early 1980s. This type of computation is based on the reflection of travelling localisations, which represent values of logical variables. It started with the implementation of functionally complete set (and thus universal computation) with gliders colliding in Conway's Game of Life cellular automaton~\cite{berlekamp2001winning}, conservative logic~\cite{fredkin1982conservative} and billiard ball model~\cite{margolus1984physics}. These ideas were further developed into prototypes of collision-based logical gates and circuits implemented with excitation wave-fragments in Belousov-Zhabotinsky medium~\cite{adamatzky2004collision, costello2011towards}, growing slime mould~\cite{adamatzky2010slime}, swarms of soldier crabs~\cite{gunji2011robust} and intra-cellular vesicles~\cite{mayne2015computing}. 

Aspiring to merge the above concepts, paradigms and implementations in a very simple, minimalist circuit, we designed, modelled and experimentally implemented a binary micro-fluidic half-adder which implements two logical gates at one junction; namely  exclusive disjunction and conjunction.

\subsection{Design Concept}

\begin{figure}[!tbp]
\centering
\subfigure[]{\includegraphics[scale=0.8]{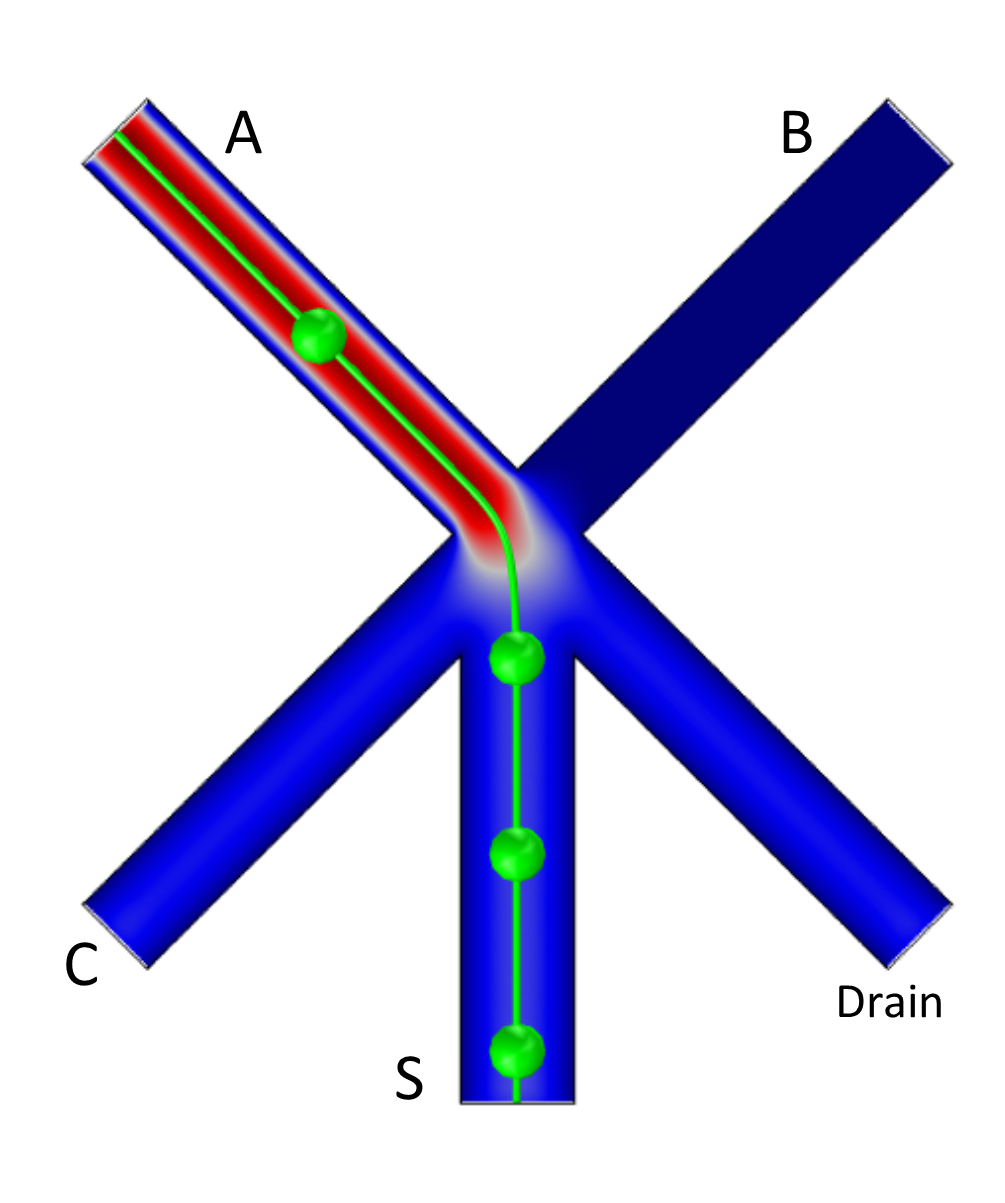}}
\subfigure[]{\includegraphics[scale=0.8]{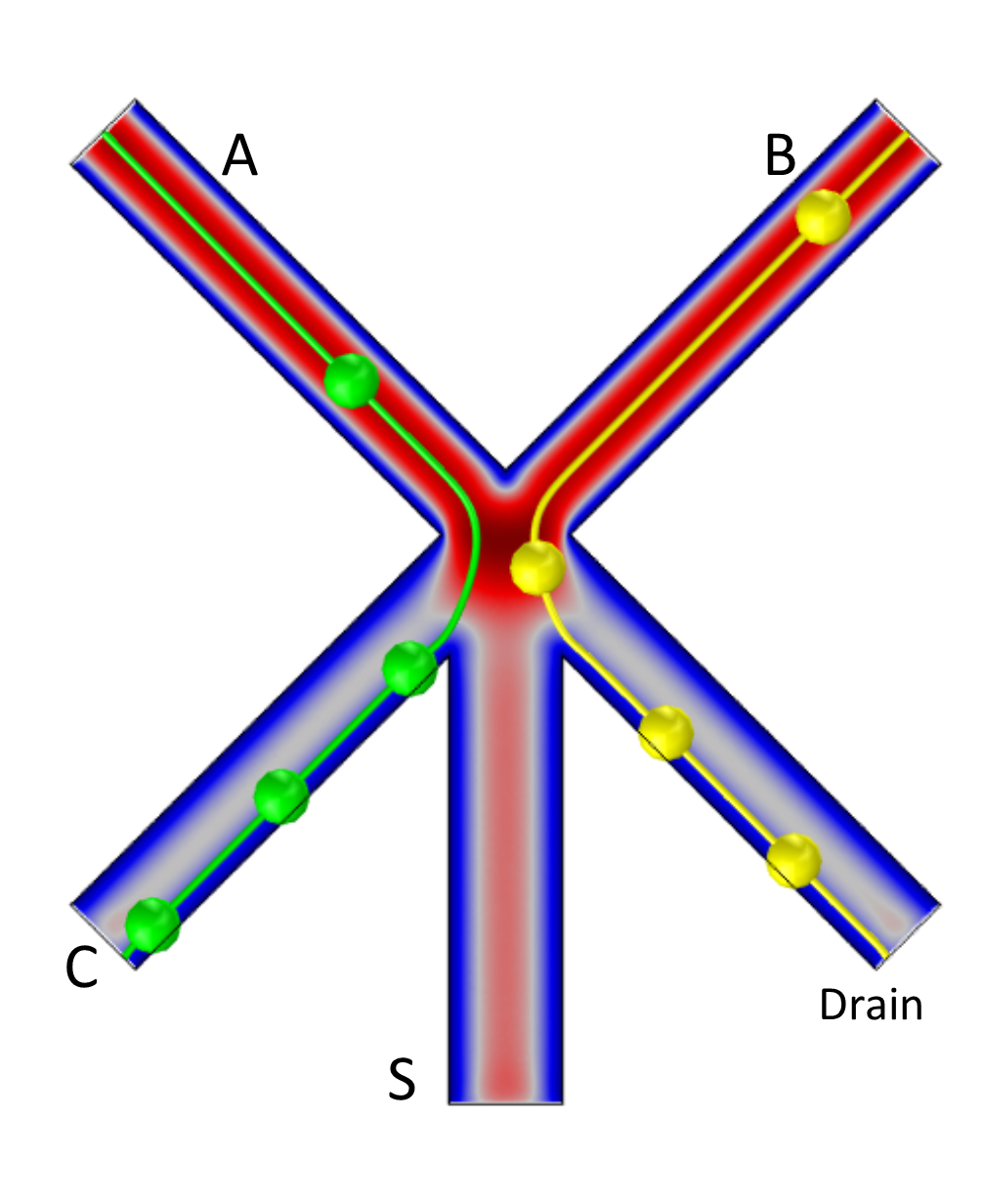}}
\caption{{\bf COMSOL Multiphysics\textsuperscript{\textregistered} simulations of the microfluidic half adder.} Stream line represents the path a droplet should take through the junction. Red areas represent high flow rate and blue represents low flow rate. Droplets travel through the central, sum, channel when a single input is {\sc True} (01, 10). When both inputs are  {\sc True} (11) the droplets are deflected by the flow such that they travel down the carry and drain channels.
(a) Inputs: $A=1$, $B=0$; Outputs: $S=1$, $C=0$. 
(b) Inputs: $A=1$,  $B=1$; Outputs: $S=0$,  $C=1$. 
}
\label{fig2}
\end{figure}

The half-adder presented here is based on the principle of fluidic resistance. The flow of fluids through a tube or channel is analogous to that of current along a wire and, as such, there is an equivalent to Ohm's law. In electronics, Ohm's law states that voltage is equal to current multiplied by resistance. Similarly in fluidics, the pressure differential along the channel, $\Delta$\textit{P}, is equal to the volumetric flow rate, \textit{Q}, multiplied by the fluidic resistance, \textit{R}. 
\begin{equation}
\Delta P = QR
\end{equation}
When considering continuous laminar flow of incompressible liquids in a rectangular channel, the Hagen-Poiseiulle equation can be adapted to give:
\begin{equation}
\Delta P = \frac{32U_{avg}\mu L}{D_h^2}
\end{equation}
where $U_{avg}$ is the average flow velocity of the fluid, $\mu$\ is the fluid viscosity, $L$ is the channel length and $D_h^2$ is the hydraulic diameter given by:
\begin{equation}
D_h=\frac{4A}{p}
\end{equation}
with $A$ being the cross-sectional area of the channel and $p$ the wetting perimeter of the channel. Using these equations, it is possible to calculate the fluidic resistance as:
\begin{equation}
R=\frac{32\mu L}{D_h^2A}
\end{equation}
By designing channels of appropriate lengths and diameters it is possible to predict the path a droplet will take through a network of fluidic channels, that being the path of least resistance. The half-adder presented here uses this principle to direct droplets to a given outlet, either by allowing them to travel through the path of least resistance or by actively deflecting them to a different path. The design of the half-adder can be seen in Fig.~\ref{fig2}. The droplets arriving at the junction from one of the inputs will preferentially travel down the central channel, representing the sum channel, unless there is also flow from the other input channel. In this case the droplets will travel down the side channels; the 
carry and drain channels. Droplets that go to the drain outlet are ignored.

\section{Materials and Methods}
\subsection{Microfluidic Device}

Microfluidic devices were fabricated by machining 50mm diameter PTFE discs using CNC milling (30,000 rpm, 8mm/s). The PTFE discs were placed within a metal manifold beneath a 100$\mu$m thick FEP film and a glass window. Finger tight connectors were used to attach tubing to the underside of the manifold and 12 screws were used to apply a compressive force across the glass window to ensure that the fluidic channels were sealed. All channels were milled to a depth of 0.8mm. The sum channel was 1mm wide whilst all other channels were 0.8mm wide. Olive Oil and water were delivered to the device via tubing connected to syringes and driven by syringe pumps. Food colouring was used to dye the water and Oil Red O to dye the oil for illustrative purposes. A  {\sc True} or '1' state was created by turning on flow at an input; a {\sc False} or '0' state given by no flow at the input. Droplets were created at a T-junction prior to entering the half-adder junction as indicators of the state.

\subsection{Analysis}

Flow simulations were performed using COMSOL Multiphysics\textsuperscript{\textregistered} using the 'Creeping Flow' physics module, which solves the Navier Stokes equations for conservation of momentum. Due to the complexity of simulating droplets, requiring large amounts of time or computing power for even simple analysis, the flow was simulated without droplets and streamlines were used to determine the expected path of a droplet. A   {\sc True} (or '1') input was represented by an input pressure of 500Pa whilst a {\sc False} (or '0') was simulated as a dead end due to the incompressible nature of fluids. The 'Particle Tracing for Fluid Flow' module was used to illustrate droplets moving through the device.

\section{Results}

\subsection{Analysis}

Initial simulations were undertaken to confirm that flow would be redirected such that the junction would behave as a half adder. The results of these simulations can be seen in Fig.~\ref{fig2}, with a single streamline used to represent the path a droplet would take through the junction. If only one of the inputs is {\sc True} then droplets propagate into the central output channel (Fig.~\ref{fig2}a).  If both inputs are {\sc True} then droplets travelling along the input channels are deflected at the junction (Fig.~\ref{fig2}b). One droplet enters the left branch and another droplet enters the drain (right branch).  These simulations show that output channel $S$ implements exclusive disjunction of input variables: $A \oplus B$ and output channel $C$ implements conjunction of input variables $A\cdot B$. That is, the architecture (Fig.~\ref{fig2}) implements a binary one-bit half-adder.

\subsection{Experimental}

\begin{figure}[!tbp]
\centering
\subfigure[]{\includegraphics[scale=0.8]{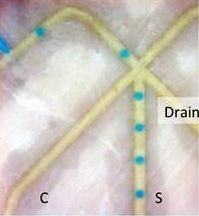}}
\subfigure[]{\includegraphics[scale=0.8]{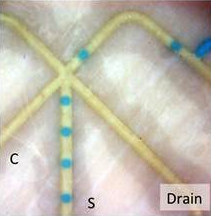}}
\subfigure[]{\includegraphics[scale=0.7]{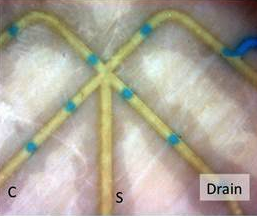}}
\subfigure[]{\includegraphics[scale=0.7]{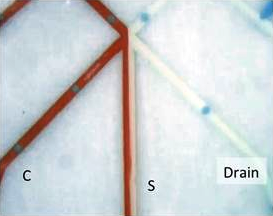}}
\caption{{\bf Microfluidic half adder in operation}. Oil flow rate 16ml/hr, Water flow rate 1ml/hr.
(a) Inputs: $A=1$, $B=0$; Outputs: $S=1$, $C=0$. 
(b) Inputs: $A=0$, $B=1$; Outputs: $S=1$, $C=0$. 
(c ) Inputs: $A=1$, $B=1$; Outputs: $S=0$, $C=1$. 
(d) Deflection of flows for inputs $A=1$ and $B=1$: Oil from input $A$ is dyed with Oil Red O. 
}
\label{fig3}
\end{figure}

The half adder was tested at a number of different flow rates. The oil flow rate was tested at 8ml/hr and 16ml/hr and the water flow rate was 0.4ml/hr and 1ml/hr (8:0.4, 16:0.4 and 16:1). The junction was found to work for all three flow rate profiles. The results for 16:1 flow rates can be seen in Fig.~\ref{fig3}.

\section{Discussion}

\begin{figure}[!tbp]
\centering
\subfigure[]{\includegraphics[scale=0.51]{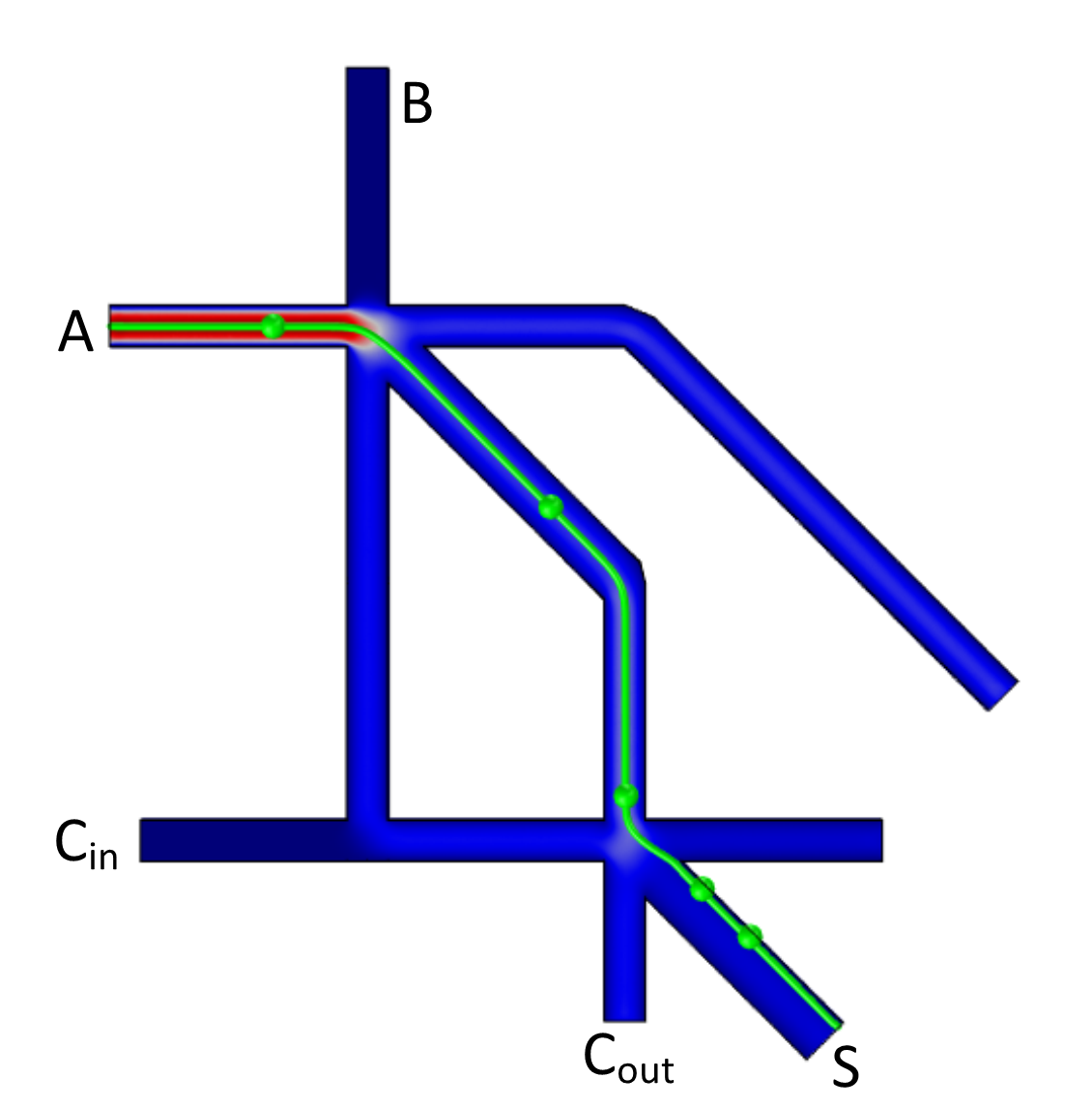}}
\subfigure[]{\includegraphics[scale=0.51]{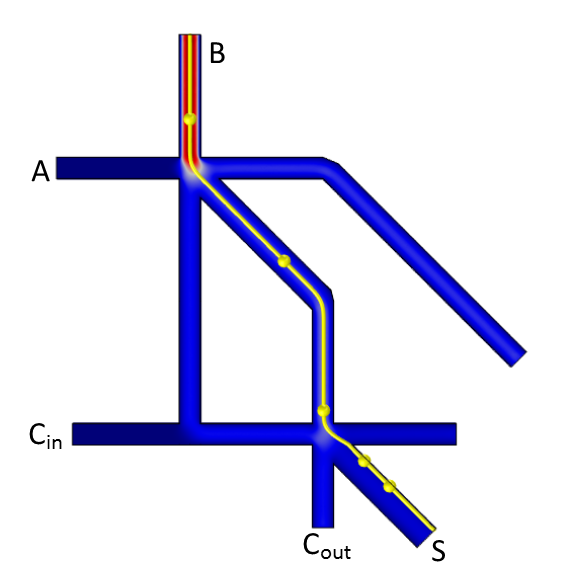}}
\subfigure[]{\includegraphics[scale=0.51]{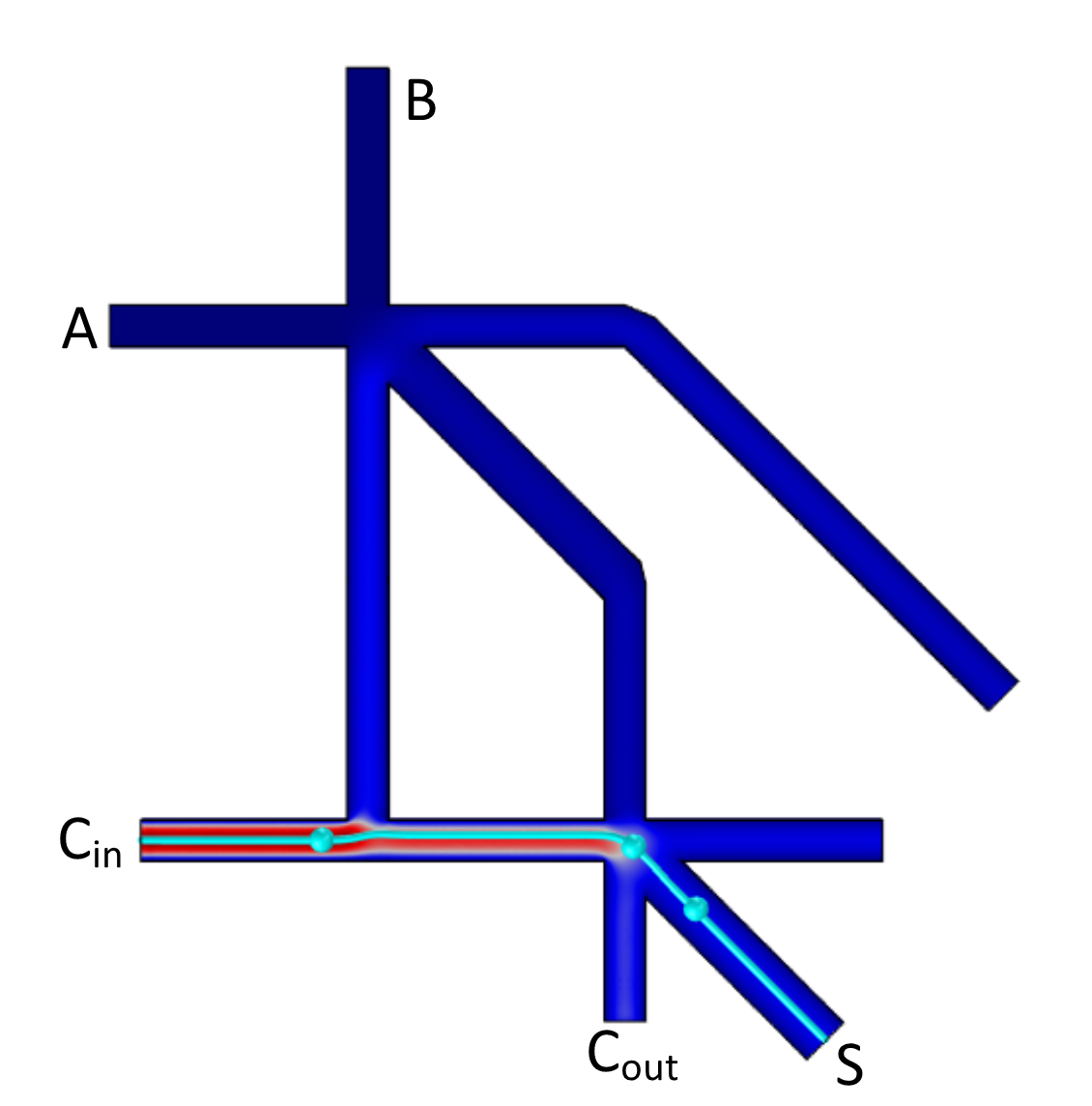}}
\subfigure[]{\includegraphics[scale=0.51]{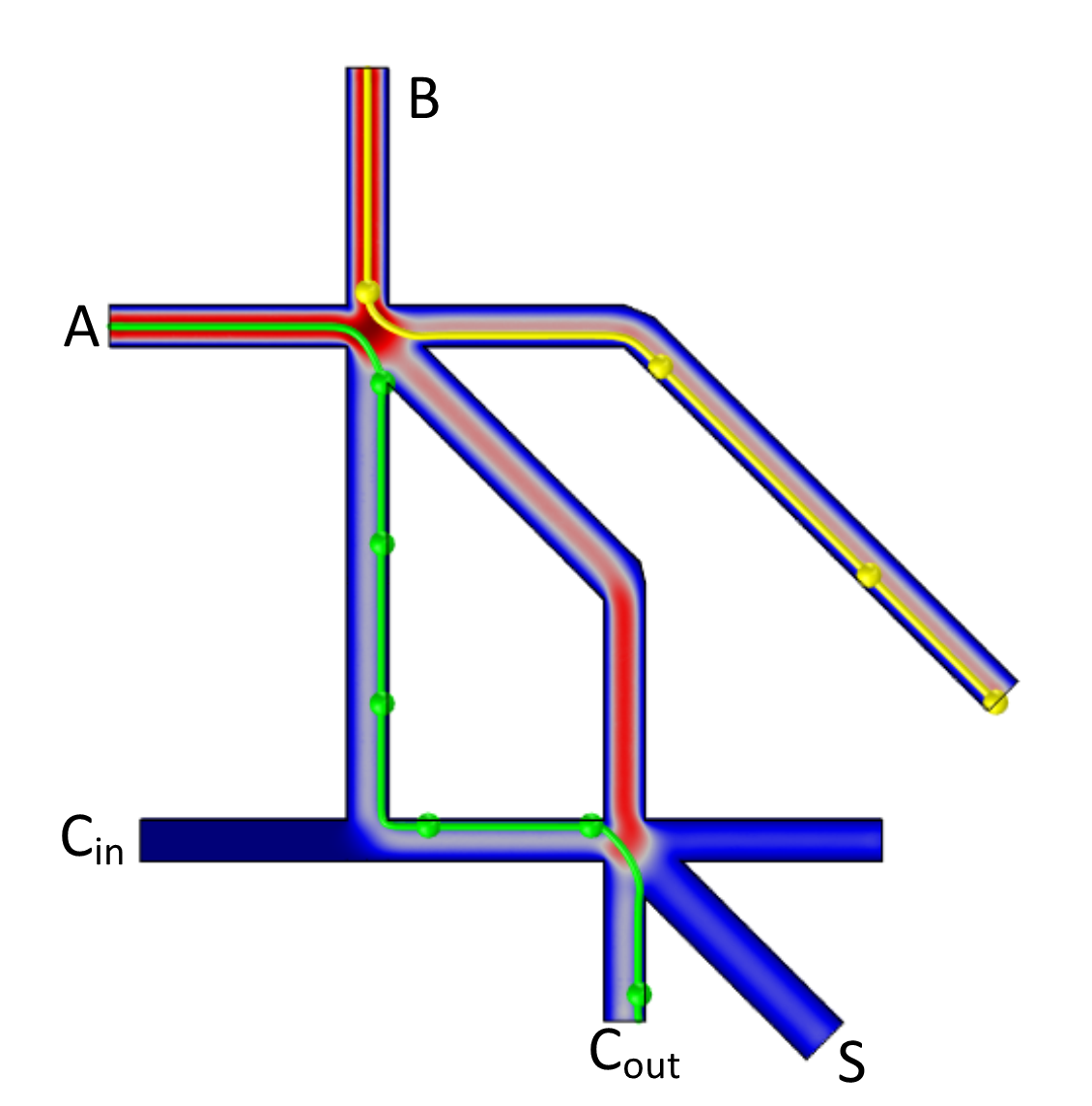}}
\subfigure[]{\includegraphics[scale=0.51]{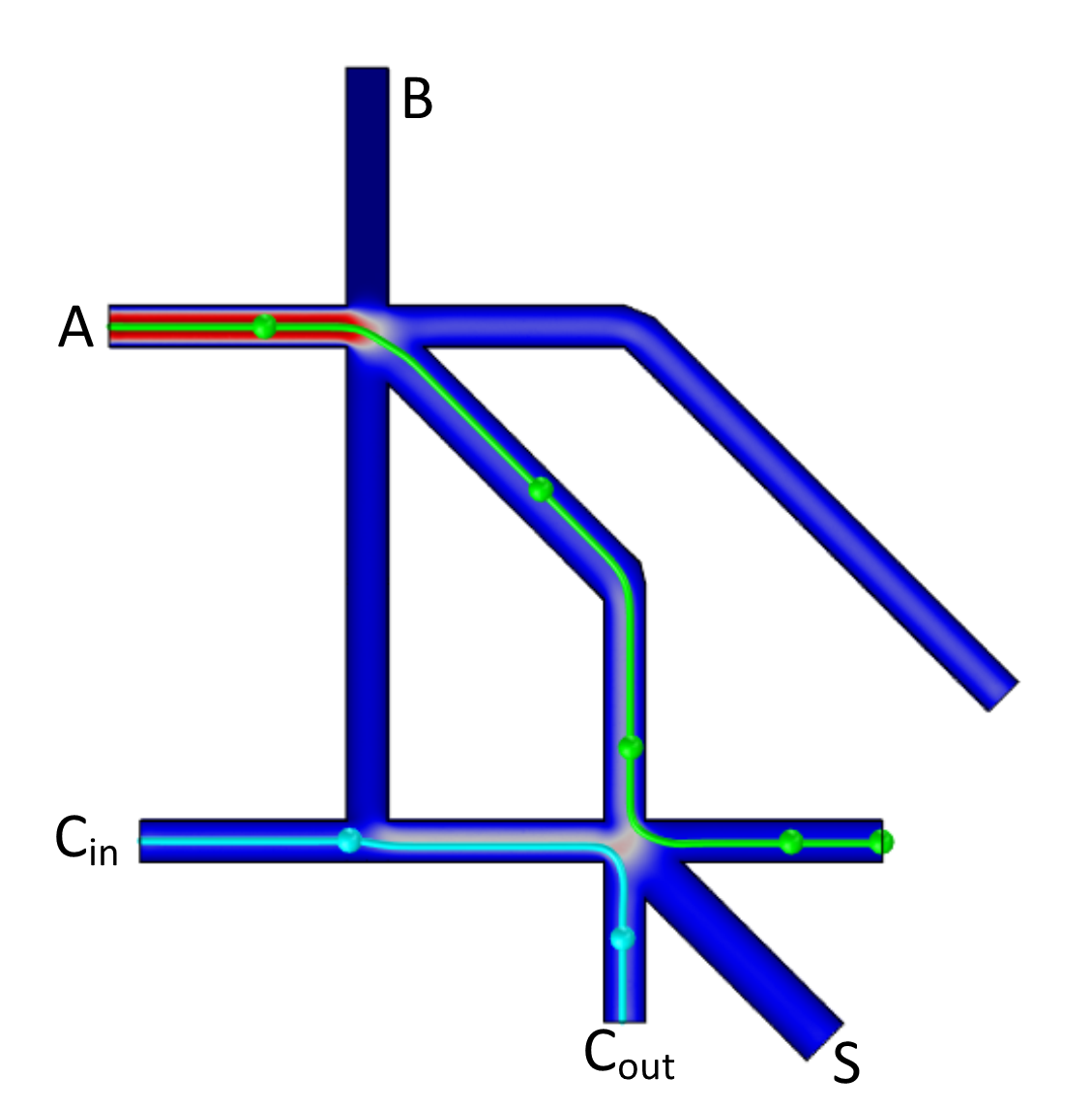}}
\subfigure[]{\includegraphics[scale=0.51]{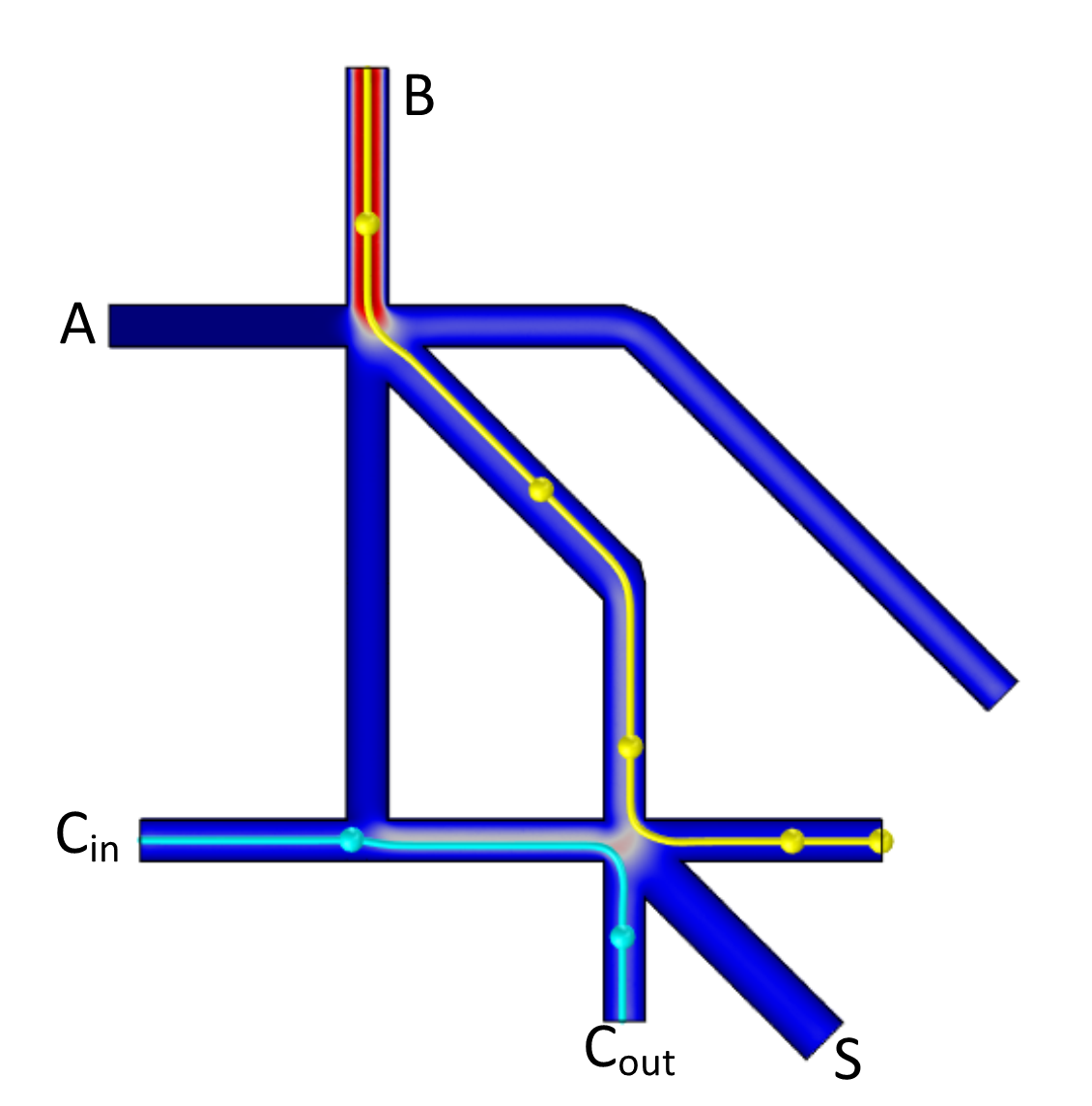}}
\subfigure[]{\includegraphics[scale=0.51]{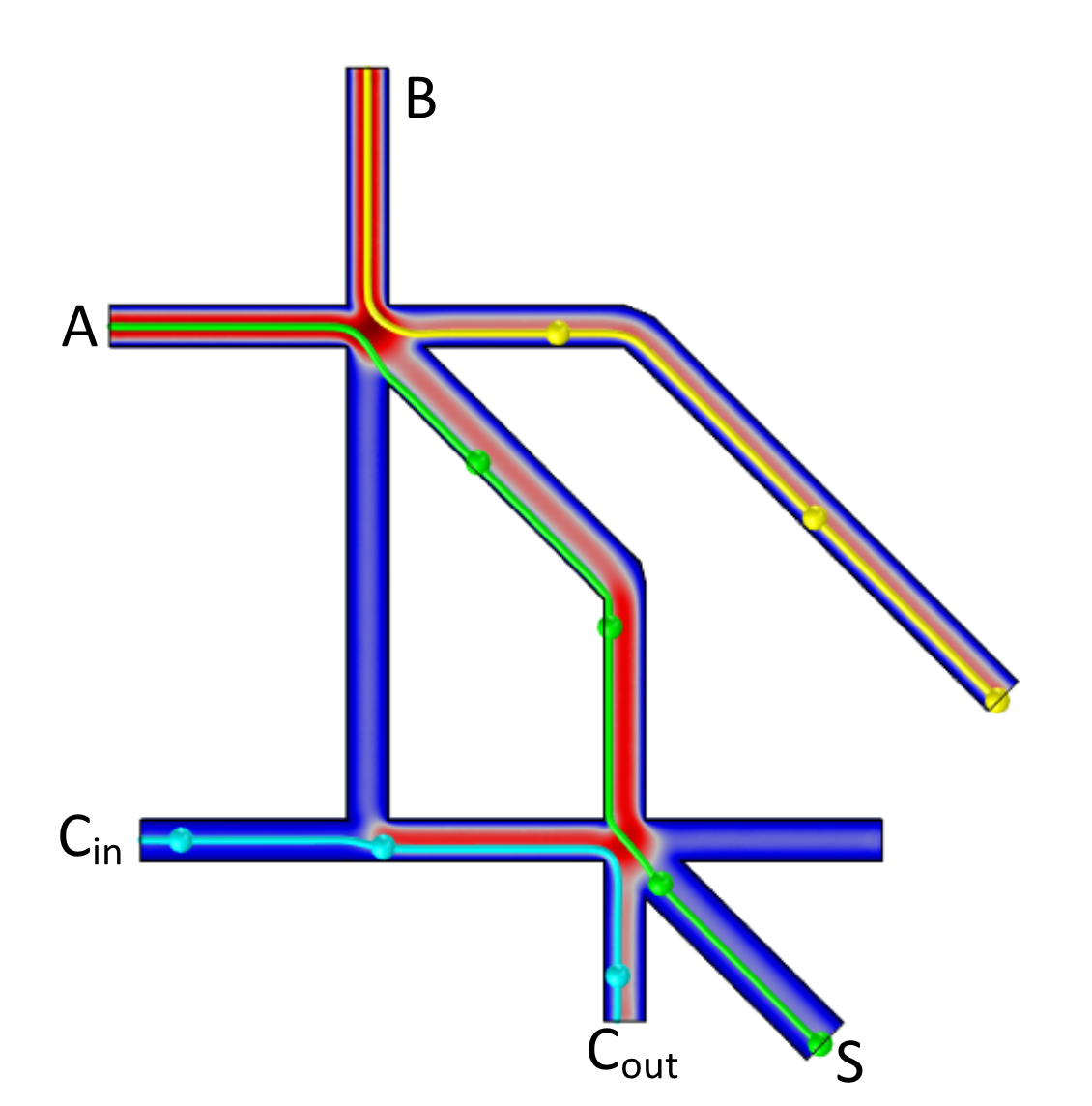}}
\caption{{\bf COMSOL Multiphysics\textsuperscript{\textregistered} simulations of the microfluidic full adder.}. 
(a) $A=1$, $B=0$, $C_{in}=0$, 
(b) $A=0$, $B=1$, $C_{in}=0$. 
(c ) $A=0$, $B=0$, $C_{in}=1$,
(d) $A=1$, $B=1$, $C_{in}=0$,
(e) $A=1$, $B=0$, $C_{in}=1$, 
(f) $A=0$, $B=1$, $C_{in}=1$, 
(g) $A=1$, $B=1$, $C_{in}=1$.Stream lines indicate the expected flow direction of a droplet assuming no collisions with other droplets. Pressure applied to inlets, pressure at inlet $C_{in}$ is half that of inlets $A$ and $B$. Red indicates faster flow, dark blue slower flow (darkest blue is stationary flow).
}
\label{fulladdersimulation}
\end{figure}

With regard to the functional expansions of the proposed design, the half-adder described above can be cascaded to a full one-bit adder. The function of the full adder is shown in simulation in Fig.~\ref{fulladdersimulation}, where the trajectories of droplets, representing {\sc True}, are also indicated. When only one of the inputs $A$, $B$ or $C_{in}$ is {\sc True}, and other inputs are {\sc False}, a pressure is applied to the fluid in corresponding input channel and the droplet moves along the path of least resistance towards an output channel $S$ (Fig.~\ref{fulladdersimulation}abc). If both inputs $A$ and $B$ are {\sc True} but $C_{in}$ is {\sc False}, the droplets originating in $B$ deflect into the drain channel (rightmost output channel in 
 Fig.~\ref{fulladdersimulation}d), and the droplets from $A$ move towards the left channel before entering the second half adder junction where it is then deflected into the
 $C_{out}$ channel. To ensure that $A$ exits through $C_{out}$ this channel must be short enough for the fluidic resistance of $C_{out}$ and $S$ to be approximately equal. Droplet $A$ ($B$) deflects to the drain channel when only one of $A$ or $B$ is {\sc True} and $C_{in}$ is {\sc True} (Fig.~\ref{fulladdersimulation}ef). The fluid streams behave similarly at sequential half adder junctions as in the independent half-adder junctions. $A$ ($B$) exit through the central junction at the first adder before interacting with fluid stream $C_{in}$ at the second junction deflecting $C_{in}$ to $C_{out}$ and $A$ ($B$) to drain. When all three inputs are {\sc True} (Fig.~\ref{fulladdersimulation}g) the presence of flow at $C_{in}$ is sufficient to deflect $A$ through the central channel at the first junction. This then causes droplets from $C_{in}$ and $A$ to exit through $C_{out}$ and $S$ respectively.   Thus the output channel $C_{out}$ implements $A\cdot B + C\cdot (A \oplus B)$ and the output channel $S$ implements $A \oplus B \oplus C$.  Note, that droplets exit via one of two drain channels (rightmost outputs in Fig.~\ref{fulladdersimulation}) for the following combinations of inputs $ABC_{in}$: 110 (Fig.~\ref{fulladdersimulation}d), 101 (Fig.~\ref{fulladdersimulation}e),  011 (Fig.~\ref{fulladdersimulation}f), 111 (Fig.~\ref{fulladdersimulation}g).  This means that the drain output implements disjunction of pairwise conjunctions of input variables $A\cdot B+A\cdot C+B\cdot C$. The ability to cascade the junction to form a full adder presents a key potential advantage over similarly simple existing fluidic logic systems such as the one presented by Vestad et al ~\cite{vestad2004flow}.
 
 With regard to practical implementation, our designs suffer from lack of 'autonomy', as external forces are used to generate and control streams and the overall system behavior. An improvement of the design includes `passivisation' of the device, for example,  by using capillary action to move fluids without any external force applied~\cite{kim2005passive}. Another direction of experimental studies will be realisation of many-valued logical gates using divergent droplets labelled with chemical or physical identity, e.g. electrical charges~\cite{hendricks1962charged, krohn1963glycerol}, that may selectively influence the interaction of droplets at junctions, e.g. droplets with similar charge will repel and droplets with differing charge will attract at short distances. Use of self-propulsion droplets~\cite{horibe2011mode}, or chemotactic droplets~\cite{cejkova2014dynamics},  interacting with fluid flows~\cite{velarde1998drops} and with other droplets at longer distances might lead to development of hierarchical architectures of programmable fluidic-droplet systems. Alternative mechanism, e.g. fusion of droplets, by analogy with excitation wave fusion gates~\cite{adamatzky2015binary}, could also be explored. 
 
\section{Conclusion}

A simple microfluidic half-adder junction based on fluidic resistance has been demonstrated both through simulation and experimentation. The input and output states were indicated by the presence of droplets. It has been shown that the same outputs are achieved across a range of flow rates. Simulations were carried out to demonstrate potential for cascading the simple half adder junction to produce a full adder.

\section*{Acknowledgments}

The research was partially supported by the Leverhulme Trust Visiting Professorship to Martin Hanczyc.

\bibliography{bibdropletlogic} 
\bibliographystyle{plain} 

\end{document}